\documentclass[letterpaper,twocolumn,fleqn]{article} 

\usepackage{ist}

\usepackage{multirow}    %
\usepackage{booktabs}    %
\usepackage{array}       %
\usepackage{multirow}    %
\usepackage{graphicx}    %
\usepackage{adjustbox}   %
\usepackage{caption}     %
\usepackage{xcolor}
\usepackage{hyperref}

\title{PySpatial: A High-Speed Whole Slide Image Pathomics Toolkit}

\author{ 
Yuechen Yang, Department of Computer Science, Vanderbilt University, Nashville, TN \\
Yu Wang, Department of Biostatistics, Vanderbilt University, Nashville, TN \\
Tianyuan Yao, Department of Computer Science, Vanderbilt University, Nashville, TN \\
Ruining Deng, Department of Computer Science, Vanderbilt University, Nashville, TN \\
Mengmeng Yin, Department of Pathology, Microbiology and Immunology, Vanderbilt University, Nashville, TN \\
Shilin Zhao, Department of Biostatistics, Vanderbilt University, Nashville, TN \\
Haichun Yang, Department of Pathology, Microbiology and Immunology, Vanderbilt University, Nashville, TN \\
Yuankai Huo, Department of Computer Science, Vanderbilt University, Nashville, TN}

\date{} 

\begin{document} 

\maketitle 

\thispagestyle{empty} 

\begin{figure*}[ht]
    \centering
    \includegraphics[width=0.9\textwidth]{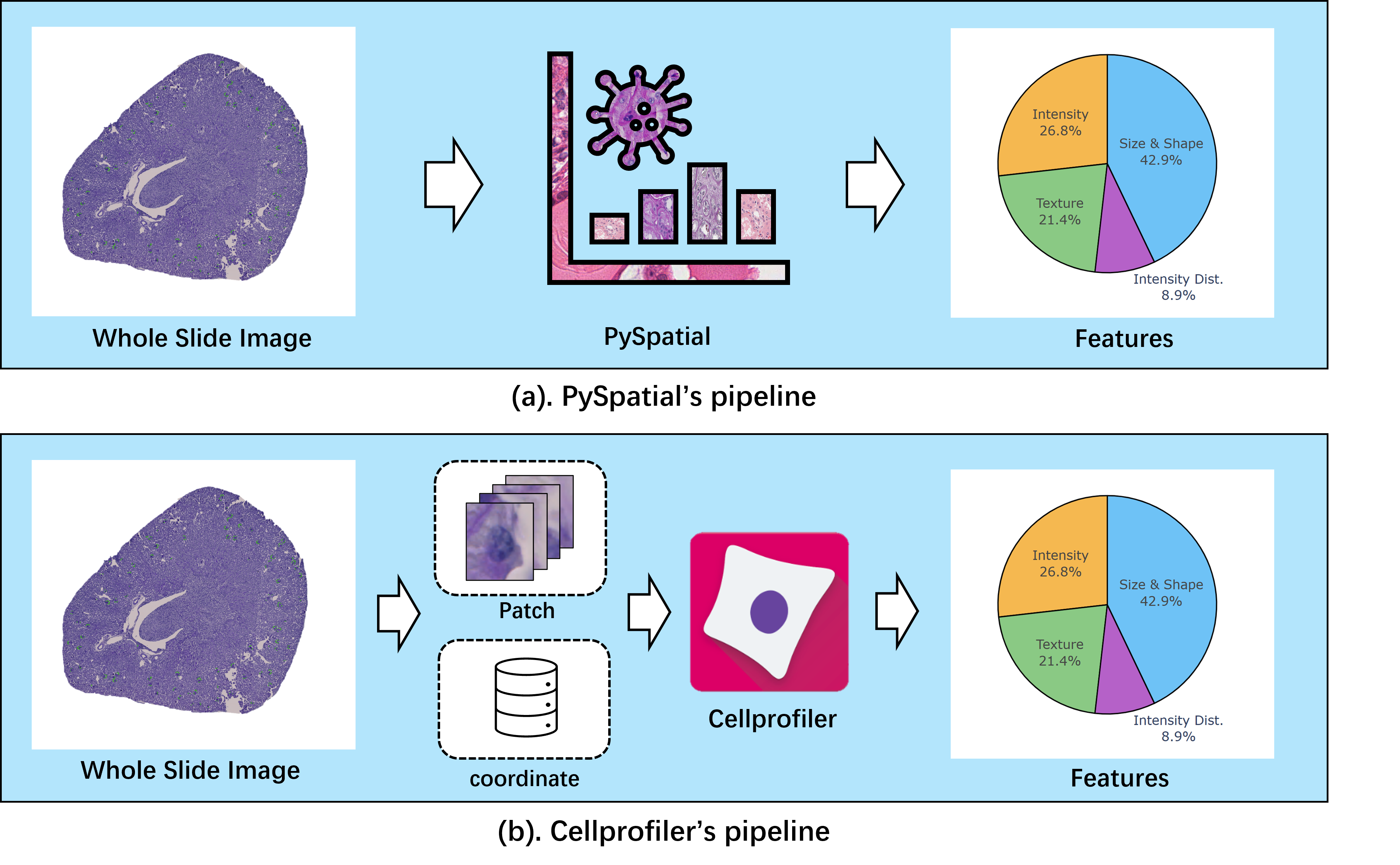} 
    \caption{Comparison between PySpatial and CellProfiler pipelines for WSI feature extraction. The top panel illustrates the PySpatial pipeline, which directly processes Whole Slide Images (WSIs) to extract features, simplifying the workflow and improving efficiency. In contrast, the bottom panel represents the traditional CellProfiler pipeline, where WSIs are first divided into smaller patches, with their coordinates recorded, followed by feature extraction at the patch level, and finally mapping features back to the original WSI. PySpatial eliminates the intermediate patch-level processing, resulting in a more streamlined and computationally efficient workflow.}
    \label{fig:pipeline}
\end{figure*}


\begin{abstract}
{Whole Slide Image (WSI) analysis plays a crucial role in modern digital pathology, enabling large-scale feature extraction from tissue samples\cite{gupta2019emergence}. However, traditional feature extraction pipelines based on tools like CellProfiler\cite{stirling2021cellprofiler} often involve lengthy workflows, requiring WSI segmentation into patches, feature extraction at the patch level, and subsequent mapping back to the original WSI\cite{chen2024spatial}. To address these challenges, we present PySpatial, a high-speed pathomics toolkit specifically designed for WSI-level analysis. PySpatial streamlines the conventional pipeline by directly operating on computational regions of interest, reducing redundant processing steps. Utilizing rtree-based spatial indexing and matrix-based computation, PySpatial efficiently maps and processes computational regions, significantly accelerating feature extraction while maintaining high accuracy. Our experiments on two datasets—Perivascular Epithelioid Cell (PEC) and data from the Kidney Precision Medicine Project (KPMP)~\cite{kpmp}—demonstrate substantial performance improvements. For smaller and sparse objects in PEC datasets, PySpatial achieves nearly a 10-fold speedup compared to standard CellProfiler pipelines. For larger objects, such as glomeruli and arteries in KPMP datasets, PySpatial achieves a 2-fold speedup. These results highlight PySpatial’s potential to handle large-scale WSI analysis with enhanced efficiency and accuracy, paving the way for broader applications in digital pathology.
}
\end{abstract}


\section{Introduction}
\label{sec:intro}

Whole Slide Imaging (WSI) has revolutionized digital pathology, providing high-resolution, comprehensive visualizations of entire tissue slides~\cite{kothari2013pathology}. These large-scale images enable pathologists to analyze intricate tissue structures and cellular features, supporting both diagnostic and research workflows\cite{gilley2024utilizing}\cite{holscher2023next}\cite{fogo2023learning}. Despite its advantages, the inherent size and complexity of WSIs pose significant computational challenges, particularly in feature extraction and quantitative analysis\cite{chakroun2018gpu}.

Existing tools, such as CellProfiler \cite{stirling2021cellprofiler}, have been instrumental in advancing image-based analysis in pathology. However, these tools typically operate at the patch level, and the workflow based on these tools often involve multi-step workflows that include segmenting WSIs into smaller patches, extracting features from each patch, and subsequently mapping these features back to the original WSI\cite{chen2024spatial}. While effective, these workflows are computationally expensive, time-consuming, and introduce significant overhead due to redundant processing steps\cite{chakroun2018gpu}.

To address these limitations, we introduce PySpatial, a high-speed pathomics toolkit specifically designed for WSI-level analysis. PySpatial significantly simplifies the traditional pipeline by eliminating the need for intermediate patch-level segmentation and mapping. Instead, it operates directly on the WSI, reducing unnecessary processing steps and streamlining the entire workflow. This difference is illustrated in Figure~\ref{fig:pipeline}, where PySpatial's pipeline (top panel) directly processes WSIs to extract features, bypassing the patch segmentation and coordinate mapping stages required by CellProfiler (bottom panel). By avoiding these intermediate steps, PySpatial achieves a more efficient and scalable pipeline, particularly for large-scale WSI datasets.

Additionally, PySpatial leverages rtree-based spatial indexing to efficiently establish relationships between computational regions and the original WSI. Furthermore, matrix-based batch computation accelerates feature extraction by performing operations across entire computational regions simultaneously, rather than on individual objects sequentially.

We validated PySpatial using two datasets: Perivascular Epithelioid Cell (PEC) and Kidney Precision Medicine Project (KPMP)\cite{kpmp}. The PEC dataset, characterized by small and sparse objects, demonstrated a nearly 10-fold speedup compared to traditional workflows. In contrast, the KPMP dataset, containing larger objects such as non-globally-sclerotic glomeruli, globally-sclerotic glomeruli, and arteries/arterioles, achieved a 2-fold speedup. These results highlight PySpatial's robustness and efficiency across diverse data types.

The key innovations and contributions of PySpatial are as follows:

\begin{itemize}
\item \textbf{Direct WSI-Level Pathomics Analysis}: PySpatial enables feature extraction directly at the WSI level, eliminating the need for patch-based segmentation and significantly shortening the conventional processing pipeline.

\item \textbf{Improved Feature Extraction Speed}: By focusing on computationally relevant regions and utilizes matrix-level operations, PySpatial accelerates feature extraction by approximately 10-fold speedup on small objects (PEC dataset) and nearly double speedup on larger objects (the KPMP dataset)  compared to existing methods.

\end{itemize}

\begin{figure*}[ht]
    \centering
    \includegraphics[width=0.9\textwidth]{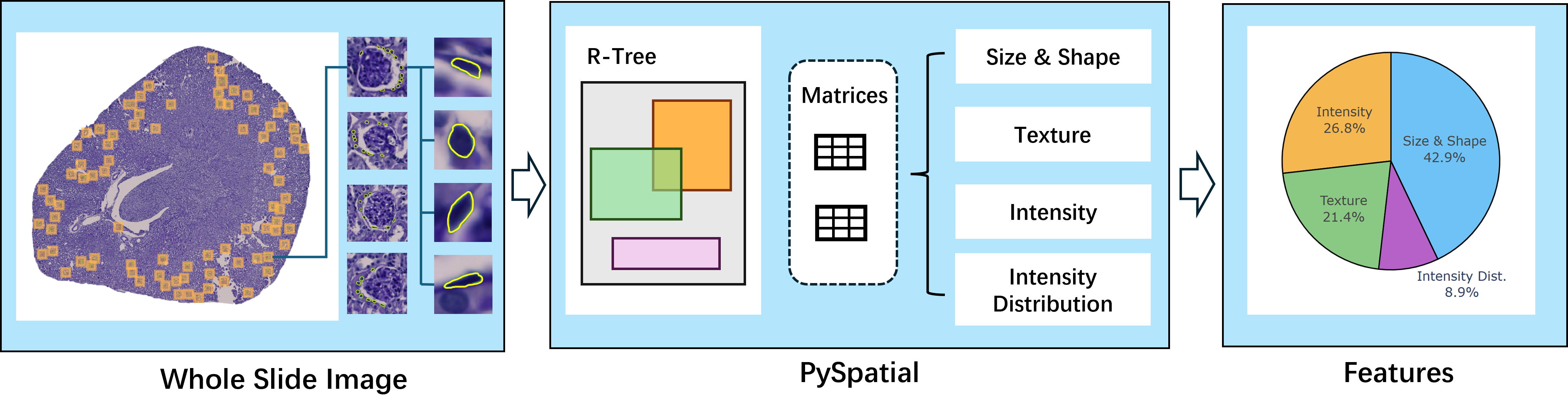}
    \caption{PySpatial's computational workflow for WSI feature extraction. The figure illustrates the computational workflow of PySpatial for feature extraction from Whole Slide Images (WSIs). In the first stage, the computational regions (highlighted in orange) are identified within the WSI, focusing only on areas of interest for analysis. In the second stage, an R-tree spatial index is built to efficiently map and manage these computational regions, which are then matrix-encoded for batch processing. Four key feature categories are extracted using specialized modules: Size \& Shape, Texture, Intensity, and Intensity Distribution. In the last graph, the output features are summarized and visualized in a pie chart, representing the proportion of each feature type extracted from the WSI.}
    \label{fig:workflow}
\end{figure*}

\section{Methods}

\subsection{PySpatial Workflow Details}
As shown in Figure~\ref{fig:workflow},  the PySpatial workflow consists of several key stages, including the extraction of computational regions, efficient matrix-based batch computation, and comprehensive feature extraction, all while maintaining spatial relationships through an R-tree mapping structure~\cite{rslan2017spatial}.

Whole Slide Images (WSIs) are typically labeled by deep learning models or manual annotation by pathologists to define regions of interest (ROIs). These labeled regions, referred to as Computational Regions, represent areas containing meaningful pathological information. However, these regions are often sparse and scattered across the WSI, as shown in Figure~\ref{fig:workflow}, where the orange-highlighted areas represent labeled computational regions in a PEC dataset. This sparsity underscores the need for a focused and computationally efficient approach to feature extraction.

To preserve the spatial context of these computational regions, PySpatial employs an R-tree spatial indexing structure\cite{zhang2007new}. This structure establishes an efficient mapping between the extracted computational regions and their original spatial locations on the WSI. The R-tree not only facilitates rapid access and retrieval of specific computational regions but also ensures that the spatial relationships between objects remain intact throughout the processing workflow.

Once the computational regions are identified and indexed, PySpatial converts these regions into a matrix representation, enabling high-throughput batch processing. Instead of analyzing individual objects sequentially, PySpatial performs operations across the entire matrix simultaneously. For instance, geometric properties such as perimeter and area can be calculated in a single batch operation across all objects within the computational regions. This matrix-based computation approach eliminates redundant calculations, reduces computational overhead, and significantly accelerates the feature extraction process.

Following the matrix computation, PySpatial extracts features across four primary categories: Size \& Shape, Texture, Intensity, and Intensity Distribution. The Size \& Shape module captures geometric attributes such as area, perimeter, and object morphology. The Texture module analyzes surface patterns and texture granularity within the objects. The Intensity module measures pixel intensity values, reflecting image brightness and contrast. Lastly, the Intensity Distribution module evaluates how pixel intensities are spatially distributed across the computational regions. Together, these modules generate a 247-dimensional feature vector for each object, providing a rich and comprehensive representation of the WSI's pathomic characteristics.

After feature extraction, PySpatial uses the R-tree index to map the extracted features back onto the original WSI. This step ensures that each feature is spatially aligned with its corresponding object, preserving the spatial context for downstream analysis and visualization. This final mapping step bridges the gap between computational processing and spatial interpretation, allowing pathologists and researchers to link extracted features with their original spatial locations on the WSI.

Through this streamlined workflow, PySpatial combines computational region focus, R-tree spatial mapping, matrix-based batch computation, and modular feature extraction to achieve efficient, scalable, and spatially accurate pathomics analysis at the WSI level.

\begin{figure}[ht]
    \centering
    \includegraphics[width=0.9\columnwidth]{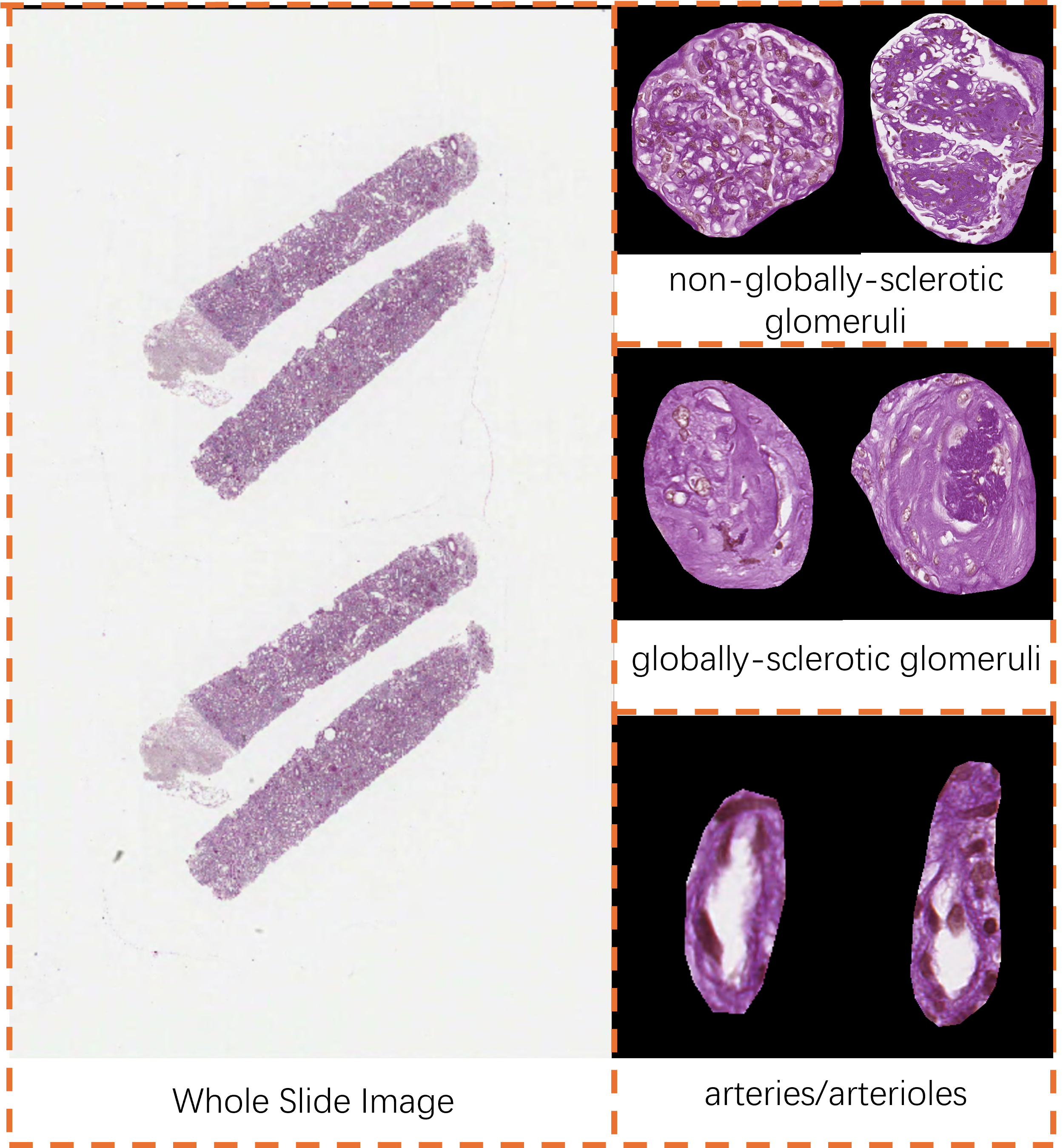}
    \caption{Overview of KPMP Dataset Object Categories.The image illustrates representative examples of the three object categories in the KPMP dataset: non-globally-sclerotic glomeruli (top), globally-sclerotic glomeruli (middle), and arteries/arterioles (bottom). These objects are characterized by their large size and relatively sparse distribution within WSIs.}
    \label{fig:KPMP_data}
\end{figure}


\section{Experiments}

In this section, we describe the datasets, experimental setup, and specific considerations related to memory optimization in PySpatial. Two datasets, PEC and KPMP, were employed to evaluate the performance and scalability of PySpatial compared to the traditional CellProfiler workflow. The experimental design focuses on analyzing pipeline efficiency, feature extraction accuracy, and the computational challenges posed by large objects in WSIs.

\subsection{Datasets}
Two distinct datasets, PEC and KPMP, were used to evaluate the PySpatial workflow.

The PEC dataset was manually annotated by pathologists using the QuPath software, a widely adopted tool for digital pathology annotation~\cite{humphries2021qupath}\cite{bankhead2017qupath}. The annotated regions of interest (ROIs) were then exported as geojson files, which record the spatial location and geometry of each object within the WSIs. A key characteristic of the PEC dataset is that the objects are small in size but numerous within a single WSI. This high object density creates a computational challenge when processing large-scale WSIs efficiently.

The KPMP dataset\cite{kpmp} (Kidney Precision Medicine Project) was directly provided with pre-defined mask annotations that identify three distinct object types: non-globally-sclerotic glomeruli, globally-sclerotic glomeruli, and arteries/arterioles, as shown in Figure~\ref{fig:KPMP_data}. In contrast to the PEC dataset, the KPMP objects are larger and fewer in number within each WSI. These larger objects introduce unique computational challenges, particularly when attempting to process them in batch matrix operations due to memory limitations.

The contrasting characteristics of these datasets—small and densely distributed objects in PEC versus large and sparsely distributed objects in KPMP—offer complementary scenarios for evaluating PySpatial's adaptability and efficiency across different pathological image analysis tasks.

\subsection{Experimental Setup}

To benchmark PySpatial against traditional workflows, both datasets were processed using two distinct pipelines: the traditional CellProfiler workflow and the PySpatial workflow.

In the CellProfiler workflow, the WSI is first divided into smaller patches, and the spatial coordinates of each patch are recorded. Each patch is then independently analyzed to extract object-level features. Finally, the extracted features are mapped back to their original spatial locations in the WSI using the recorded coordinates. While this approach is functional, the multiple intermediate steps (e.g., patching, coordinate mapping, and feature merging) introduce computational overhead and increase processing time.

In contrast, the PySpatial workflow eliminates the intermediate patching step by directly focusing on computational regions. These regions are defined based on annotations from either deep learning models or manual labeling by pathologists. Using an R-tree spatial index, PySpatial establishes a mapping between the extracted computational regions and their spatial locations in the original WSI. The computational regions are then converted into matrix representations, enabling efficient batch processing for feature extraction. Once the feature extraction process is complete, the R-tree index ensures that the features are accurately mapped back onto the WSI for downstream analysis.

For each dataset, the processing time for both workflows was recorded and compared to assess PySpatial's speed-up performance. Additionally, the accuracy of extracted features was evaluated to ensure consistency between PySpatial and CellProfiler results.

\subsection{Memory Constraints and Matrix Optimization}

While PySpatial's matrix-based batch computation significantly accelerates feature extraction, memory constraints can arise when processing datasets with large objects such as those found in the KPMP dataset. Each computational region is matrix-encoded for batch processing, and the size of these matrices is directly influenced by the dimensions and number of objects within a region.

In the case of KPMP data, where individual objects are large but relatively few, the matrix size can exceed system memory limits, making batch computation infeasible for certain regions. To address this issue, PySpatial allows users to set a customizable matrix size parameter, enabling them to optimize memory usage based on their hardware capabilities.

Furthermore, when matrix-based computation is no longer efficient or feasible due to exceptionally large objects, PySpatial provides an alternative API that allows for direct computation on individual objects without matrix aggregation. This flexibility ensures that PySpatial remains robust and adaptable across diverse datasets, regardless of object size or distribution.

\section{Results}

\subsection{Feature Description}

\begin{figure}[htbp]
    \centering
    \includegraphics[width=0.8\columnwidth]{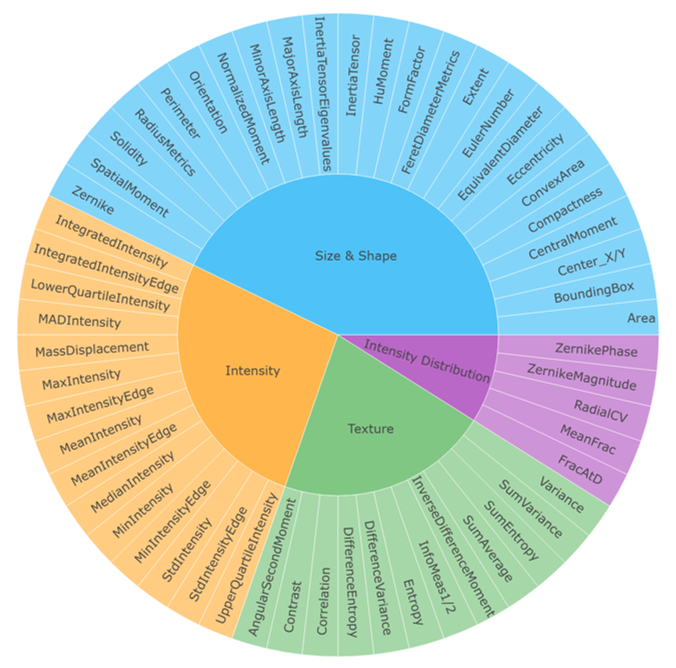}
    \caption{Overview of Feature Categories and Subtypes Extracted by PySpatial. The diagram illustrates the 247-dimensional feature space categorized into Size \& Shape, Texture, Intensity, and Intensity Distribution, with representative subtypes shown in the outer ring.}
    \label{fig:fig4}
\end{figure}

PySpatial extracts a comprehensive set of 247 features categorized into four primary groups: Size \& Shape, Texture, Intensity, and Intensity Distribution. As shown in Figure~\ref{fig:fig4}, each category encompasses multiple quantitative descriptors. For example, Size \& Shape includes features like Area and Perimeter, Texture captures patterns through metrics such as Entropy and Contrast, Intensity reflects pixel-level properties like Mean Intensity and Max Intensity, while Intensity Distribution examines spatial intensity patterns through descriptors like Zernike Phase and Radial CV. This rich feature set enables PySpatial to provide a detailed and multi-dimensional representation of objects within WSIs, supporting downstream analysis and interpretation.

\begin{figure*}[ht]
    \centering
    \includegraphics[width=0.9\textwidth]{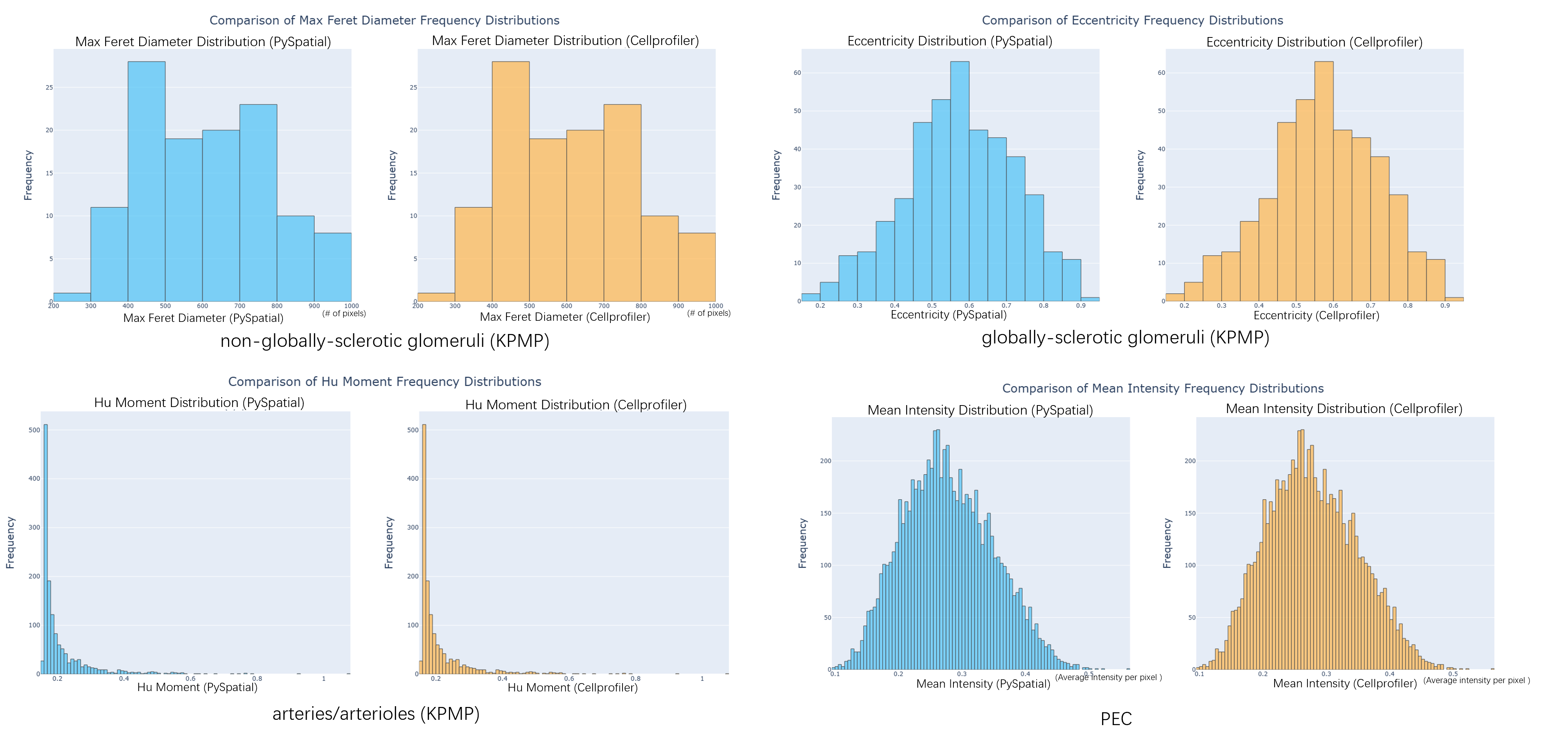}
    \caption{ Comparison of Feature Frequency Distributions between PySpatial and CellProfiler.
The figure presents frequency distributions of selected features (Max Feret Diameter, Eccentricity, Hu Moment, and Mean Intensity) extracted by PySpatial (left panels) and CellProfiler (right panels). Across different object types, both tools exhibit highly consistent distribution patterns, demonstrating the accuracy and reliability of PySpatial's feature extraction results.}
    \label{fig:result}
\end{figure*}

\subsection{Running Performance Comparison}
\begin{figure}[ht]
    \centering
    \includegraphics[width=0.9\columnwidth]{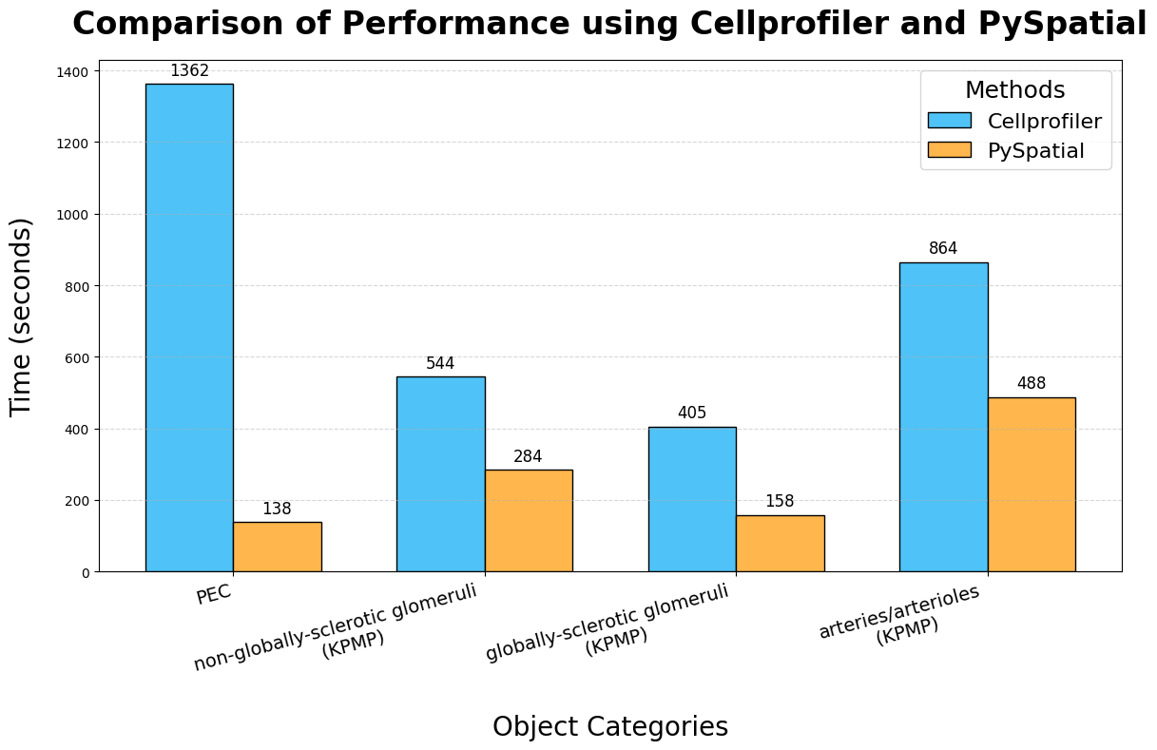}
    \caption{Comparison of processing time between CellProfiler and PySpatial across different datasets. The bar chart compares the performance of CellProfiler (blue bars) and PySpatial (orange bars) in terms of processing time across four categories: PEC, non-globally-sclerotic glomeruli, globally-sclerotic glomeruli, and arteries/arterioles. PySpatial demonstrates a significant reduction in processing time across all categories, particularly in PEC data, where the improvement is most pronounced. These results highlight PySpatial's efficiency in handling both small, densely distributed objects (PEC) and larger, sparsely distributed objects (KPMP categories).}
    \label{fig:comparison}
\end{figure}

The performance comparison between PySpatial and the traditional CellProfiler workflow in terms of processing time is shown in Figure~\ref{fig:comparison}. This comparison covers two datasets: PEC and KPMP, where the latter includes three object categories: non-globally-sclerotic glomeruli, globally-sclerotic glomeruli, and arteries/arterioles. The results demonstrate a consistent and substantial reduction in processing time when using PySpatial across all categories.

In the PEC dataset, characterized by numerous small and densely distributed objects, PySpatial achieved a nearly 10-fold speedup compared to CellProfiler. This significant improvement highlights the efficiency of PySpatial's matrix-based batch processing approach, which minimizes redundant calculations and optimizes computational operations across small objects.

In contrast, the KPMP dataset, where objects are larger and relatively sparse, still exhibited notable performance gains with PySpatial. Across the three categories—non-globally-sclerotic glomeruli, globally-sclerotic glomeruli, and arteries/arterioles—PySpatial achieved approximately a 2-fold speedup on average. While the acceleration in the KPMP dataset is less pronounced compared to the PEC dataset, it underscores PySpatial's flexibility and scalability in managing varying object sizes and distributions.

The difference in speed improvement between the two datasets can be primarily attributed to object characteristics. The PEC dataset benefits extensively from batch matrix computation, which efficiently processes a large number of small objects simultaneously. On the other hand, the KPMP dataset's larger objects sometimes encounter memory constraints during matrixization, limiting the full potential of batch computation. However, PySpatial also provides API, which allows users to skip matrixization and directly compute object-level features, provides an effective workaround to these constraints.

Overall, PySpatial consistently outperformed CellProfiler in terms of processing time, regardless of object size or distribution. The results demonstrate that PySpatial not only accelerates feature extraction but also adapts effectively to varying computational challenges posed by different datasets. This flexibility makes PySpatial a robust and efficient toolkit for large-scale WSI-level pathomics analysis.

\subsection{Consistency Between PySpatial and CellProfiler}
To validate the consistency of feature extraction between PySpatial and CellProfiler, we compared the frequency distributions of several representative features across datasets, including Max Feret Diameter, Eccentricity, Hu Moment, and Mean Intensity. The results are illustrated in Figure~\ref{fig:result}.

The distributions of these features extracted by PySpatial (left panels) and CellProfiler (right panels) exhibit remarkably similar patterns, with matching peak positions, range distributions, and overall shapes. 

The observed consistency not only validates PySpatial's computational accuracy but also demonstrates its reliability for WSI-level pathomics feature extraction. 

\section{Conclusion}

In this work, we introduced PySpatial, a high-speed whole slide image (WSI) pathomics toolkit designed to address the limitations of traditional feature extraction workflows. Built upon the robust foundation of CellProfiler, PySpatial significantly simplifies the WSI analysis pipeline by eliminating intermediate steps such as patch-level segmentation and coordinate mapping. Instead, PySpatial operates directly on computational regions using R-tree spatial indexing and matrix-based batch computation, enabling efficient and scalable feature extraction at the WSI level.

Through extensive evaluation on two datasets—PEC and KPMP—we demonstrated PySpatial's remarkable performance improvements. In the PEC dataset, characterized by small and densely distributed objects, PySpatial achieved nearly a 10-fold speedup compared to CellProfiler. In the KPMP dataset, where objects are larger and more sparsely distributed, PySpatial maintained a consistent 2-fold speedup.

Furthermore, we validated the accuracy and consistency of the features extracted by PySpatial. By comparing area feature frequency distributions from both workflows, we proved that PySpatial produces results that are highly consistent with CellProfiler, ensuring computational accuracy while maintaining significant efficiency gains.

In conclusion, PySpatial represents a robust, efficient, and scalable solution for WSI-level pathomics analysis. Its ability to balance computational speed and feature extraction accuracy makes it a valuable toolkit for large-scale digital pathology studies. Future work will focus on expanding PySpatial's capabilities, including support for additional feature categories and further optimization for extremely large WSI datasets.

\section{Acknowledgment}
This research was supported by NIH R01DK135597 (Huo), DoD HT9425-23-1-0003 (HCY), and KPMP Glue Grant. This work was also supported by Vanderbilt Seed Success Grant, Vanderbilt Discovery Grant, and VISE Seed Grant. This project was supported by The Leona M. and Harry B. Helmsley Charitable Trust grant G-1903-03793 and G-2103-05128. This research was also supported by NIH grants R01EB033385, R01DK132338, REB017230, R01MH125931, and NSF 2040462. We extend gratitude to NVIDIA for their support by means of the NVIDIA hardware grant. This work was also supported by NSF NAIRR Pilot Award NAIRR240055.

The KPMP is funded by the following grants from the NIDDK: U01DK133081, U01DK133091, U01DK133092, U01DK133093, U01DK133095, U01DK133097, U01DK114866, U01DK114908, U01DK133090, U01DK133113, U01DK133766, U01DK133768, U01DK114907, U01DK114920, U01DK114923, U01DK114933, U24DK114886, UH3DK114926, UH3DK114861, UH3DK114915, UH3DK114937. The content is solely the responsibility of the authors and does not necessarily represent the official views of the National Institutes of Health.



\begin{biography}
Yuechen Yang is a first-year PhD student in Computer Science Department at Vanderbilt University. Her research focuses on medical image processing, with a particular emphasis on the extraction and application of pathomics.
\end{biography}


\small
\begin{thebibliography}{9}


\bibitem{gupta2019emergence} R. Gupta, T. Kurc, A. Sharma, J. S. Almeida, and J. Saltz, "The emergence of pathomics," \emph{Current Pathobiology Reports}, vol. 7, pp. 73–84, 2019, Springer.

\bibitem{stirling2021cellprofiler} D. R. Stirling, M. J. Swain-Bowden, A. M. Lucas, A. E. Carpenter, B. A. Cimini, and A. Goodman, "CellProfiler 4: improvements in speed, utility and usability," \emph{BMC Bioinformatics}, vol. 22, pp. 1–11, 2021, Springer.


\bibitem{holscher2023next} D. L. Hölscher, N. Bouteldja, M. Joodaki, M. L. Russo, Y.-C. Lan, A. V. Sadr, M. Cheng, V. Tesar, S. V. Stillfried, B. M. Klinkhammer, et al., "Next-Generation Morphometry for pathomics-data mining in histopathology," \emph{Nature Communications}, vol. 14, no. 1, p. 470, 2023, Nature Publishing Group UK London.



\bibitem{chen2024spatial} J. Chen, Y. Wang, R. Deng, Q. Liu, C. Cui, T. Yao, Y. Liu, J. Zhong, A. B. Fogo, H. Yang, et al., "Spatial pathomics toolkit for quantitative analysis of podocyte nuclei with histology and spatial transcriptomics data in renal pathology," in \emph{Medical Imaging 2024: Digital and Computational Pathology}, vol. 12933, pp. 252–260, 2024, SPIE.


\bibitem{gilley2024utilizing} P. Gilley, K. Zhang, N. Abdoli, Y. Sadri, L. Adhikari, K.-M. Fung, and Y. Qiu, "Utilizing a pathomics biomarker to predict the effectiveness of bevacizumab in ovarian cancer treatment," \emph{Bioengineering}, vol. 11, no. 7, p. 678, 2024.


\bibitem{humphries2021qupath} M. P. Humphries, P. Maxwell, and M. Salto-Tellez, "QuPath: The global impact of an open source digital pathology system," \emph{Computational and Structural Biotechnology Journal}, vol. 19, pp. 852–859, 2021, Elsevier.

\bibitem{bankhead2017qupath} P. Bankhead, M. B. Loughrey, J. A. Fernández, Y. Dombrowski, D. G. McArt, P. D. Dunne, S. McQuaid, R. T. Gray, L. J. Murray, H. G. Coleman, et al., "QuPath: Open source software for digital pathology image analysis," \emph{Scientific Reports}, vol. 7, no. 1, pp. 1–7, 2017, Nature Publishing Group.


\bibitem{kothari2013pathology} S. Kothari, J. H. Phan, T. H. Stokes, and M. D. Wang, "Pathology imaging informatics for quantitative analysis of whole-slide images," \emph{Journal of the American Medical Informatics Association}, vol. 20, no. 6, pp. 1099–1108, 2013, BMJ Publishing Group.


\bibitem{chakroun2018gpu} I. Chakroun, N. Michiels, and R. Wuyts, "GPU-accelerated CellProfiler," in \emph{2018 IEEE International Conference on Bioinformatics and Biomedicine (BIBM)}, pp. 321–326, 2018, IEEE.



\bibitem{rslan2017spatial} E. Rslan, H. A. Hameed, and E. Ezzat, "Spatial R-tree index based on grid division for query processing," \emph{Int. J. Database Manag. Syst. (IJDMS)}, vol. 9, no. 6, pp. 25–36, 2017.


\bibitem{zhang2007new} Z.-B. Zhang, J.-P. Zhang, J. Yang, and Y. Yang, "A new approach to creating spatial index with R-tree," in \emph{2007 International Conference on Machine Learning and Cybernetics}, vol. 5, pp. 2645–2648, 2007, IEEE.


\bibitem{fogo2023learning} A. B. Fogo, "Learning from deep learning and pathomics," \emph{Kidney International}, vol. 104, no. 6, pp. 1050–1053, 2023, Elsevier.



\bibitem{kpmp}  “Kidney precision medicine project data.” \url{https://www.kpmp.org}. Accessed Month Day, Year. Funded by the National Institute of Diabetes and Digestive and Kidney Diseases (Grant numbers: U01DK133081, U01DK133091, U01DK133092, U01DK133093, U01DK133095, U01DK133097, U01DK114866, U01DK114908, U01DK133090, U01DK133113, U01DK133766, U01DK133768, U01DK114907, U01DK114920, U01DK114923, U01DK114933, U24DK114886, UH3DK114926, UH3DK114861, UH3DK114915, UH3DK114937).

\end{thebibliography}
\end{document}